\documentclass[conference]{IEEEtran}
\IEEEoverridecommandlockouts
\usepackage{cite}
\usepackage{amsmath,amssymb,amsfonts}
\usepackage{algorithmic}
\usepackage{graphicx}
\usepackage{textcomp}
\usepackage{xcolor}
\def\BibTeX{{\rm B\kern-.05em{\sc i\kern-.025em b}\kern-.08em
    T\kern-.1667em\lower.7ex\hbox{E}\kern-.125emX}}

\usepackage{xcolor}

\usepackage{flushend}

\usepackage{setspace}

\begin{document}

\title{Orthogonal Time-Frequency Space (OTFS) Aided Media-Based Modulation System For 6G and Beyond Wireless Communications Networks

\thanks{This work was supported by TUBITAK 1001 (Grant Number: 123E513).}}

\author{Author Name}

\author{\IEEEauthorblockN{Burak Ahmet Ozden\IEEEauthorrefmark{1}\IEEEauthorrefmark{2},
Murat Kaymaz\IEEEauthorrefmark{1}\IEEEauthorrefmark{3},
Erdogan Aydin\IEEEauthorrefmark{1}, 
Emir Aslandogan\IEEEauthorrefmark{4}, Haci Ilhan\IEEEauthorrefmark{4}, \\
Ertugrul Basar\IEEEauthorrefmark{5}\IEEEauthorrefmark{6}, 
Miaowen Wen\IEEEauthorrefmark{7}, 
Marco Di Renzo\IEEEauthorrefmark{8}}
\IEEEauthorblockA{\IEEEauthorrefmark{1}Department of Electrical and Electronics Engineering, Istanbul Medeniyet University, Istanbul, Turkey.}
\IEEEauthorblockA{\IEEEauthorrefmark{2}Department of Computer Engineering, Yildiz Technical University, Istanbul, Turkey.}
\IEEEauthorblockA{\IEEEauthorrefmark{3} Ulak Haberlesme A.S, Istanbul, Turkey.}
\IEEEauthorblockA{\IEEEauthorrefmark{4}Department of Electronics and Communication Engineering, Yildiz Technical University, Istanbul, Turkey.}
\IEEEauthorblockA{\IEEEauthorrefmark{5}Department of Electrical Engineering, Tampere University, 33720 Tampere, Finland.} 
\IEEEauthorblockA{\IEEEauthorrefmark{6}Department of Electrical and Electronics Engineering, Koc University, Istanbul, Turkey.} 
\IEEEauthorblockA{\IEEEauthorrefmark{7}School of Electronic and Information Engineering, South China University of Technology, Guangzhou, China.}
\IEEEauthorblockA{\IEEEauthorrefmark{8} Laboratoire des Signaux et Syst\`emes, CentraleSup\'elec, CNRS, Universit\'e Paris-Saclay, France.}
\\  Email: bozden@yildiz.edu.tr, murat.kaymaz@ulakhaberlesme.com.tr, erdogan.aydin@medeniyet.edu.tr, emira@yildiz.edu.tr, \\ ilhanh@yildiz.edu.tr,  
ebasar@ku.edu.tr, eemwwen@scut.edu.cn, marco.di-renzo@universite-paris-saclay.fr.
}

\maketitle

\begin{abstract}

This paper proposes a new orthogonal time frequency space (OTFS)-based index modulation system called OTFS-aided media-based modulation (MBM) scheme (OTFS-MBM), which is a promising technique for high-mobility wireless communication systems. The OTFS technique transforms information into the delay-Doppler domain, providing robustness against channel variations, while the MBM system utilizes controllable radio frequency (RF) mirrors to enhance spectral efficiency. The combination of these two techniques offers improved bit error rate (BER) performance compared to conventional OTFS and OTFS-based spatial modulation (OTFS-SM) systems. The proposed system is evaluated through Monte Carlo simulations over high-mobility Rayleigh channels for various system parameters. Comparative throughput, spectral efficiency, and energy efficiency analyses are presented, and it is shown that  OTFS-MBM outperforms traditional OTFS and OTFS-SM techniques. The proposed OTFS-MBM scheme stands out as a viable solution for sixth generation (6G) and next-generation wireless networks, enabling reliable communication in dynamic wireless environments.

\end{abstract} 


\section{Introduction}
The rapid advancement of modern transportation technologies, such as high-speed trains and autonomous vehicles, has heightened the demand for reliable and seamless communication systems. High-mobility scenarios, such as vehicles moving at speeds over 500 km/h, require stable and robust communication frameworks. However, conventional modulation techniques struggle to overcome the challenges posed by time-varying wireless channels, such as Doppler effects and multipath propagation  \cite{OTFS_SURVEY}. Moreover, the growing data traffic in wireless networks demands higher data rates and greater energy efficiency \cite{SURVEY2}. Therefore, new and advanced wireless communication technologies are needed to meet these demands and requirements.

Media-based modulation (MBM) is a high-performance index modulation (IM) technique that has emerged as a promising solution for enhancing data transmission reliability and spectral efficiency. This technique utilizes the on/off states of radio frequency (RF) mirrors near the transmit antenna to create distinct channel realizations, rather than relying on traditional signal methods.  In MBM, data bits are transmitted over indices of different channel fading realizations, known as mirror activation patterns (MAPs), which are generated by the on/off states of the RF mirrors \cite{MBM3, MBM1, MBM4, MBM2, MBM5}. This approach enhances spectral efficiency, throughput, and energy efficiency, particularly in rich scattering environments.

Orthogonal time frequency space (OTFS), which was first proposed in \cite{OTFSILK}, is an innovative modulation technique designed to overcome the challenges of high-mobility communication scenarios. Unlike traditional modulation schemes, such as orthogonal frequency division multiplexing (OFDM), which modulates information in the time-frequency domain, OTFS maps data bits onto the delay-Doppler (DD) domain, providing robustness against channel impairments like Doppler shifts and multipath delays. As a result, it substantially improves performance in high-Doppler scenarios such as vehicular and high-speed train communication systems \cite{OTFSs1, OTFSs2, RISotfs}. New OTFS-based wireless communication systems that provide high performance in high mobility scenarios compared to conventional modulation techniques have become the focus of many researchers. The study of  \cite{OTFSs3} proposes an intelligent reflecting surface-based OTFS system, employing location-aided channel estimation and an iterative interference cancellation detector to improve signal coherence. The work of  \cite{OTFSs4} introduces OTFS-based index modulation multiple access, allowing multiple users to share DD resources without coordination, enhancing bit error rate and overall system efficiency. The study of \cite{OTFSs5} suggests using precoders like Zadoff-Chu, Walsh-Hadamard, and discrete cosine transform to reduce OTFS's high peak-to-average power ratio, achieving significant reduction with minimal performance loss. Also, the work of  \cite{OTFS_SM} proposes a spatial modulation (SM) scheme for multiple-input multiple-output (MIMO)-OTFS systems to enhance spectral efficiency while reducing detection complexity compared to space-time-coded OTFS (STC-OTFS).

IM transmits data bits not only in the traditional parameters of amplitude, phase, or frequency but also in the indices of system components, such as antennas, subcarriers, MAP states, spreading codes, or time slots. IM uses available resources in wireless communication systems to enhance spectral and energy efficiency \cite{RISantenna, IM1, burak2, burak3}. Integrating OTFS with IM has emerged as a promising approach for enhancing next-generation wireless communication systems. OTFS-based IM techniques have the potential to improve error performance, spectral efficiency, and energy efficiency, as well as robustness against Doppler effects and time-varying channels. Recent studies show that OTFS-based IM systems provide more efficient data transmission with lower error rates than traditional OTFS systems. In \cite{OTFS_IM1}, joint DD index modulation OTFS is proposed, activating delay or Doppler elements with distinct constellations to improve spectral efficiency. Simulation results show its superiority over traditional OTFS under various conditions. The authors of  \cite{OTFS_IM2} introduce delay-IM with OTFS and Doppler-IM with OTFS, activating blocks of delay/Doppler elements. Bit error rate (BER) upper bounds are analyzed, and two low-complexity algorithms (multi-layer joint symbol and activation pattern (MLJSAPD) detection and customized message passing (CMP) detection) are proposed, showing robustness to imperfect channel state information. In \cite{OTFS_IM3}, OTFS-aided dual-mode IM is proposed to balance reliability and spectral efficiency. It achieves lower error rates than traditional OTFS and OTFS-based IM.

This paper proposes a new wireless communication scheme called OTFS-MBM, which combines OTFS for robustness against high-mobility wireless communication environments and MBM for high spectral efficiency into a single-input multiple-output (SIMO) architecture. The contributions of this paper are as follows:

\begin{enumerate}

\item Monte Carlo simulation of the proposed OTFS-MBM system is carried out for high mobility scenarios over Rayleigh channels with DD spreads, considering the QAM method. 

\item The proposed OTFS-MBM system is compared with OTFS and OTFS-SM systems in terms of throughput, spectral efficiency, and energy saving.

\item BER performance is analyzed under various system configurations, including variations in the number of receive antennas (\(N_R\)), modulation order (\(M_q\)), and the number of RF mirrors (\(n_{RF}\)).

\item The error performance of the OTFS-MBM system is compared with that of traditional OTFS and OTFS-SM systems for different system parameters.

\end{enumerate}

This paper organizes the remaining sections as follows: Section II presents the system models of the traditional OTFS and proposed OTFS-MBM schemes. Section III offers performance analyses, including throughput, spectral efficiency, and energy saving. Section IV demonstrates simulation results. Finally, Section V concludes the paper.

The notations used in this paper are as follows. Vectors and matrices are denoted by bold lowercase and uppercase characters, respectively. The operators $(\cdot)^T$, $\|\cdot\|$, and $\otimes$ represent the transpose, Euclidean norm, and Kronecker product.

\begin{figure*}[t]
\centering{\includegraphics[width=1\textwidth]{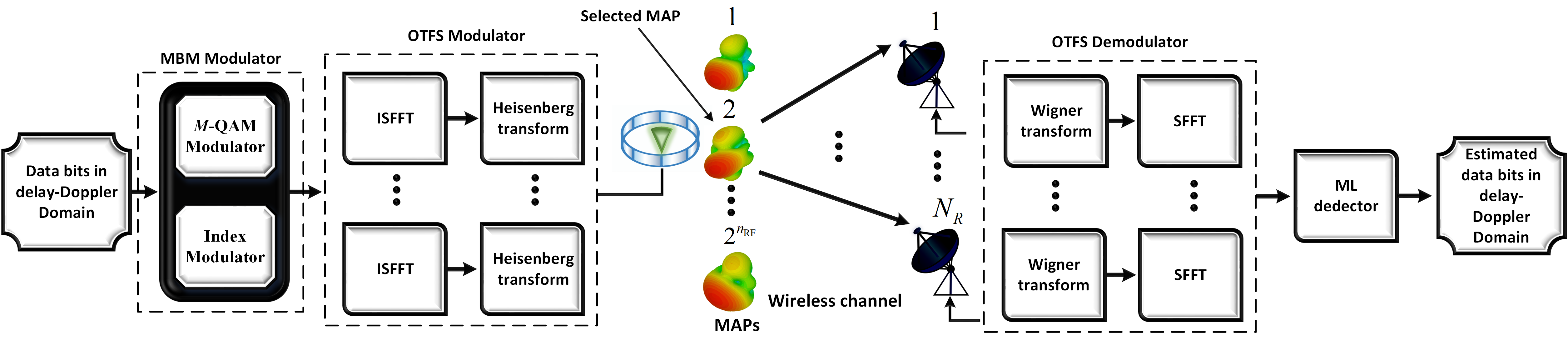}}	
	\caption{System model of the proposed OTFS-MBM scheme.}
	\label{system_model} 
\end{figure*}

\begin{table*}[t]
\centering
\addtolength{\tabcolsep}{-5.5pt}
\caption{Mapping procedure for the OTFS-MBM system}
\label{OTFS_SM_MAP}
\begin{tabular}{|c|c|c|c|c|c|} 
\hline
\hline
\textbf{\small Parameters}       & \textbf{\begin{tabular}[c]{@{}c@{}}\small Data Bits\\ \end{tabular}} & \textbf{\begin{tabular}[c]{@{}c@{}} \small First DD grid \\ \end{tabular}} & \textbf{\begin{tabular}[c]{@{}c@{}} \small Second DD grid \\ \end{tabular}} & \textbf{\begin{tabular}[c]{@{}c@{}} \small Third DD grid \\ \end{tabular}} & \textbf{\begin{tabular}[c]{@{}c@{}} \small Fourth DD grid \\ \end{tabular}} \\ \hline\hline
\begin{tabular}[c]{@{}c@{}} \small $M_q=4$\\ $n_{RF}=2$\\ $N=2$\\ $M=2$ \end{tabular} & 
\begin{tabular}[c]{@{}c@{}} {[}0111001100011101{]} \\ $\eta_{OTFS-MBM}=16$ \end{tabular}                                                             & \begin{tabular}[c]{@{}c@{}}{[}0111{]}\\ $x_p=1-j$, $\varpi=2$ \\ $\textbf{s}_1=[0 \ \ 1-j \ \ 0 \ \ 0]^T$ \end{tabular}                   & \begin{tabular}[c]{@{}c@{}}{[}0011{]} \\ $x_p=1-j$, $\varpi=1$ \\ $\textbf{s}_2=[1-j \ \ 0 \ \ 0 \ \ 0]^T$ \end{tabular}                 & \begin{tabular}[c]{@{}c@{}}{[}0001{]}\\ $x_p=-1-j$, $\varpi=1$ \\ $\textbf{s}_3=[-1-j \ \ 0 \ \ 0 \ \ 0]^T$  \end{tabular}  & \begin{tabular}[c]{@{}c@{}}{[}1101{]} \\ $x_p=-1-j$, $\varpi=4$ \\ $\textbf{s}_4=[0 \ \ 0 \ \ 0 \ \ -1-j]^T$ \end{tabular} \\ \hline \hline
\end{tabular}
\end{table*}

\section{System Model}

This section presents the conventional and proposed OTFS-MBM system models, respectively.

\subsection{Traditional OTFS System Model}
OTFS modulation is a novel wireless communication scheme designed to perform effectively in wireless environments characterized by high mobility. By utilizing the DD domain, OTFS enhances resilience to inter-symbol interference and time-varying channel effects. In an OTFS framework, $M$ subcarriers and $N$ time slots define a DD domain grid. The symbols $x[k, l]$, where $k = 0, \dots, N-1$ and $l = 0, \dots, M-1$, are selected from a modulation set $\mathcal{Q}$. The representation in the time-frequency (TF) domain is derived using the inverse symplectic finite Fourier transform (ISFFT) \cite{OTFSILK}:
\begin{equation}
X[n, m] = \frac{1}{\sqrt{NM}} \sum_{k=0}^{N-1} \sum_{l=0}^{M-1} x[k, l] e^{j2\pi\left(\frac{nk}{N} - \frac{ml}{M}\right)}.
\end{equation}
The TF symbols $X[n, m]$ are then mapped onto orthogonal subcarriers using the Heisenberg transform. The resulting time-domain signal is expressed as \cite{OTFSILK}:
\begin{equation}
s(t) = \sum_{n=0}^{N-1} \sum_{m=0}^{M-1} X[n, m] g_{\text{x}}(t - nT) e^{j2\pi m\Delta f(t-nT)},
\end{equation}
where $g_{\text{x}}(t)$ is the transmit pulse shaping function, $\Delta f$ denotes the subcarrier spacing, and $T$ represents the time slot duration.

At the receiver, the signal is processed by a matched filter, then sampled, and finally transformed back to the DD domain using the symplectic finite Fourier transform (SFFT) \cite{OTFSILK}:
\begin{equation}
y[k, l] = \frac{1}{\sqrt{NM}} \sum_{n=0}^{N-1} \sum_{m=0}^{M-1} Y[n, m] e^{-j2\pi\left(\frac{nk}{N} - \frac{ml}{M}\right)},
\end{equation}
where $Y[n, m]$ is the TF representation of the received signal.

The DD domain channel response $h( \tau, \nu)$ is modeled as the superposition of $P$ propagation paths, where each path is characterized by a delay $\tau_i$, a Doppler shift $\nu_i$, and a complex channel coefficient $h_i$ \cite{OTFSILK}:
\begin{equation}
h(\tau, \nu) = \sum_{i=1}^{P} h_i \delta(\tau - \tau_i) \delta(\nu - \nu_i).
\end{equation}
The delay $\tau_i$ and Doppler shift $\nu_i$ are normalized as:
\begin{equation}
\tau_i = \frac{k_i}{M\Delta f}, \quad \nu_i = \frac{l_i}{NT}.
\end{equation}

\subsection{The Proposed OTFS-MBM System Model}

MBM and OTFS techniques are combined in a SIMO architecture to achieve high spectral efficiency and robustness against disturbances in high mobility scenarios. This proposed new wireless communication system is named OTFS-MBM for short. Fig. \ref{system_model} presents the system model of the proposed OTFS-MBM scheme. The system model of the proposed OTFS-MBM scheme includes one reconfigurable antenna (RA) at the transmitter and $N_R$ receive antennas at the receiver. It is deployed with $n_{RF}$ RF mirrors near the RA. The on and off states of this $n_{RF}$ RF mirror generate $2^{n_{RF}}$ different MAPs. In the proposed system, these MAP indices transmit extra data in addition to the data carried in the conventional symbol. Hence, the spectral efficiency of the proposed OTFS-MBM system at each DD grid point corresponds to that of the MBM system and is expressed as follows:
\begin{equation}
    \eta_{MBM}=\log_2(M_q) + n_{RF},
\end{equation}
here, $M_q$ is the modulation order and $n_{RF}$ is the number of RF mirrors. Furthermore, for a DD grid of size $N \times M$, the spectral efficiency of the OTFS-MBM system can be formulated as follows:
\begin{equation}
    \eta_{OTFS-MBM}=NM\Big(\log_2(M_q) + n_{RF} \Big).
\end{equation}

In each MBM-OTFS transmission frame, a bit sequence vector $\mathbf{b}$ consisting of $\eta_{\text{OTFS-MBM}}$ data bits is transmitted. Firstly, the $\mathbf{b}$ vector is initially divided into $NM$ segments. Subsequently, $\log_2(M_q) + n_{RF}$ bits are allocated, where $\log_2(M_q)$ bits are mapped to the chosen symbol ($x_q$), and $n_{RF}$ bits are assigned to the active MAP index ($\varpi$). The transmission vector $\textbf{s}_i \in \mathbb{C}^{2^{n_{RF}} \times 1} $ for the OTFS-MBM system in the $i^{\text{th}}$ DD grid can be represented as follows:
\begin{eqnarray}\label{transmission_signal}
\textbf{s}_{i} = {\Bigg[ 0\, \,0\cdots 0\,  \, \cdots \underset{\stackrel{\uparrow}{\varpi^{\text{th}} \ \text{active MAP index}}}{0 \, x_q \, 0} \cdots \, \,0\, \,\cdots 0\, \,0
	\Bigg]^T},
\end{eqnarray}
where, $q \in \{1,2,\cdots, M_q\}$, $\varpi \in \{1,2,\cdots, 2^{n_{RF}}\}$, and $i \in \{1,2,\cdots, NM\}$. The transmission matrix $\textbf{S} \in \mathbb{C}^{2^{n_{RF}} \times NM}$ of the OTFS-MBM system containing all DD grid points is expressed as follows:
\begin{equation} \textbf{S}=[\textbf{s}_1, \textbf{s}_2, \cdots, \textbf{s}_{NM} ] =
\begin{bmatrix}
0 & 0 & \cdots & 0 \\
\vdots & x_q & \cdots & \vdots \\
x_q & \vdots & \cdots & 0 \\
\vdots & 0 & \cdots & x_q \\
0 & 0 & \cdots & 0
\end{bmatrix}_{ 2^{n_{RF}} \times NM}\!\!\!\!\!\!\!\!\!\!\!\!\!\!\!\!\!. 
\end{equation}
Table \ref{OTFS_SM_MAP} illustrates the mapping process of the OTFS-MBM system utilizing QAM modulation with system parameters $M_q=4$, $n_{RF}=2$, $N=2$, and $M=2$. Here, for each DD grid, the first 2 bits select the active MAP index, and the other 2 bits select the QAM symbol. Since $NM=4$, the system includes a total of $4$ DD grid points. At each of these grid points, the MBM technique is implemented. Consequently, as shown in Table \ref{OTFS_SM_MAP}, each DD grid point corresponds to a single MBM transmission vector. For the example in Table \ref{OTFS_SM_MAP}, the OTFS-MBM transmission vector can be represented as follows:
\begin{equation}
    \mathbf{S} = [\textbf{s}_{1} \textbf{s}_{2} \textbf{s}_{3} \textbf{s}_{4}] = \begin{bmatrix}
        0 & 1-j & -1-j & \ 0 \\
        \ 1-j & 0 & 0 & 0 \\
        \ 0 & 0 & 0 & 0 \\
        \ 0 & 0 & 0 & -1-j
    \end{bmatrix}.
\end{equation}
Also, the $\mathbf{S^{*}} \in \mathbb{C}^{2^{n_{RF}} NM}$ matrix is defined as follows:
\begin{equation} \mathbf{S^{*}} =
\begin{bmatrix}
\textbf{s}_1 \\
\textbf{s}_2 \\
\vdots \\
\textbf{s}_{NM}
\end{bmatrix}.
\end{equation}
The OTFS modulator in the transmitter employs the ISFFT to transform the DD domain signals, represented by vectors in $\mathbf{S^{*}}$, into TF domain signals for each MAP state. It then performs the Heisenberg transform, converting these time-frequency domain signals into time-domain signals, denoted as $\mathbf{C}(t)$. 


Furthermore, the time-varying multipath Rayleigh fading channel, $\mathbf{H}_{\text{M}}(t)$, is described as a $N_RNM \times 2^{n_{RF}}NM$-dimensional matrix, each of length $P$ (representing the number of multipath). The Rayleigh fading channel matrix $\mathbf{H}_{\text{M}}(t) \in \mathbb{C}^{N_RNM \times 2^{n_{RF}}NM}$, where unique channel delays and Doppler shifts characterize each path, is expressed as follows:
\begin{equation}
    \mathbf{H}_{\text{M}}(t) \!= \!\!
\begin{bmatrix}
    \mathbf{H}_{1,1}(t) & \mathbf{H}_{1,2}(t) & \cdots & \mathbf{H}_{1,2^{n_{RF}}}(t) \\
    \mathbf{H}_{2,1}(t) & \mathbf{H}_{2,2}(t) & \cdots & \mathbf{H}_{2,2^{n_{RF}}}(t) \\
    \vdots & \vdots & \ddots & \vdots \\
    \mathbf{H}_{N_R,1}(t) & \mathbf{H}_{N_R,2}(t) & \cdots & \mathbf{H}_{N_R, 2^{n_{RF}}}(t)
\end{bmatrix}\!\!.
\end{equation}
where $\mathbf{H}_{\varpi,z}(t)$ is the $NM \times NM$ dimensional channel matrix between $\varpi^{\text{th}}$ the MAP state and the $z^{\text{th}}$ receive antenna. Here $z \in \{1,2,\cdots, N_R\}$. 

In the transmitter terminal of the proposed OTFS-MBM system, after OTFS modulation, the time domain transmission signals are transmitted simultaneously over a wireless Rayleigh channel represented by $\mathbf{H}_{\text{M}}(t)$ using one antenna and $2^{n_{RF}}$ MAP states. The time-domain signal $\mathbf{y}(t) \in \mathbb{C}^{N_RNM \times 1}$ received at the receiver can be expressed as follows:
\begin{equation}
    \mathbf{y}(t) = \mathbf{H}_{\text{M}}(t) \otimes \mathbf{C}(t) + \mathbf{w}(t),
\end{equation}
where $\mathbf{w}(t)\in \mathbb{C}^{N_RNM \times 1}$ is the Gauss noise vector in the time domain. In the receiver terminal of the proposed OTFS-MBM system, after applying the OTFS demodulator using the Wigner transform and SFFT, the received time-domain signal is converted into the corresponding received DD domain signal $\bar{\mathbf{y}}$. The received DD domain signal $\bar{\mathbf{y}}$ is expressed as follows:
\begin{align}
\bar{\mathbf{y}} &= \left( \mathbf{F}_N \otimes \mathbf{I}_{N_RM} \right) \mathbf{H}_{\text{M}}(t) \left( \mathbf{F}^{T}_N \otimes \mathbf{I}_{2^{n_{RF}}M} \right) \mathbf{S}^{*}\nonumber \\
&+ \left( \mathbf{F}_N \otimes \mathbf{I}_{N_RM} \right) \mathbf{w}(t) = \mathbf{H}_{\text{M}}^{\text{eff}} \mathbf{S}^{*} + \mathbf{w}^{\text{eff}}
\end{align}
where the matrix $\mathbf{I}_{N_RM}$ is an identity matrix of size $N_RM$, and $\mathbf{I}_{2^{n_{RF}}M}$ is an identity matrix of size $2^{n_{RF}}M$. Additionally, $\mathbf{F}_N$ represents the discrete Fourier transform (DFT) matrix of size $N$.

\begin{figure}[t]
\centering{\includegraphics[width=0.4\textwidth]{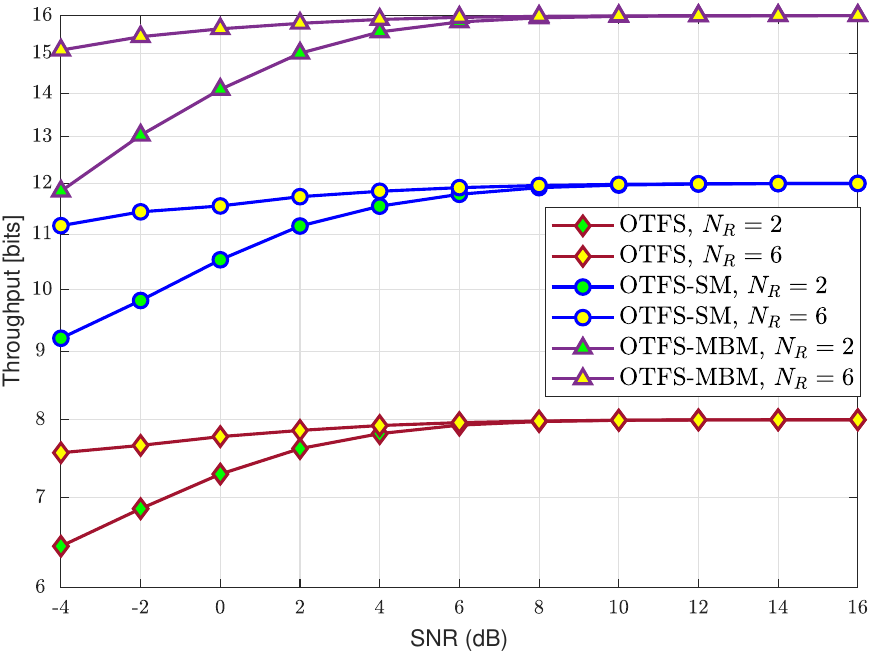}}
	\caption{Throughput comparisons of the proposed OTFS-MBM, OTFS-SM, and OTFS systems.} 
	\label{thro_mbm} 
\end{figure}

\begin{figure}[t]
\centering{\includegraphics[width=0.4\textwidth]{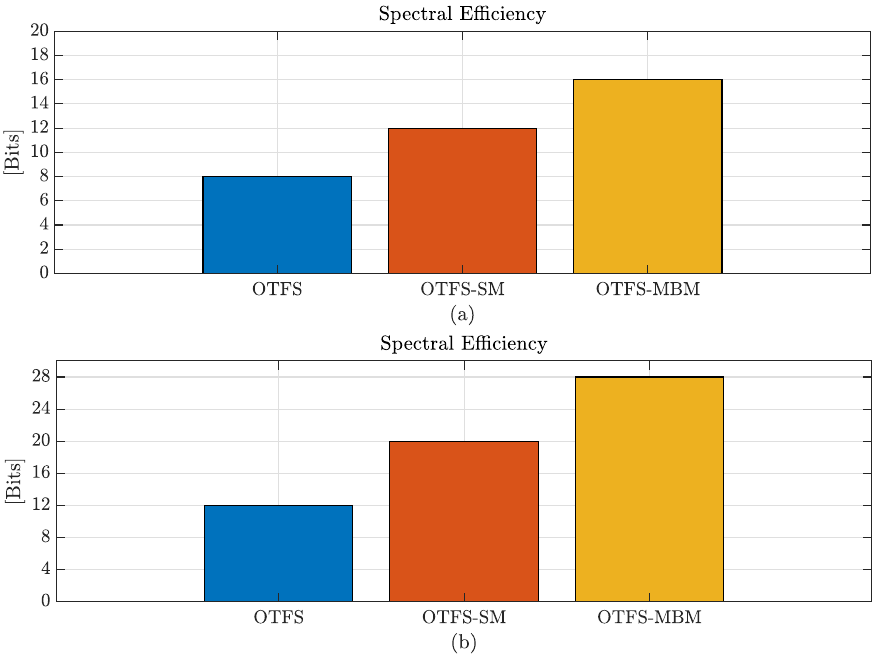}}
	\caption{Spectral efficiency comparisons of the proposed OTFS-MBM, OTFS-SM, and OTFS systems for (a) $M_q=4$, $N_T=2$, $n_{RF}=2$ and (b) $M_q=8$, $N_T=4$, $n_{RF}=4$.} 
	\label{spec_mbm} 
\end{figure}

\begin{figure}[t]
\centering{\includegraphics[scale=0.45]{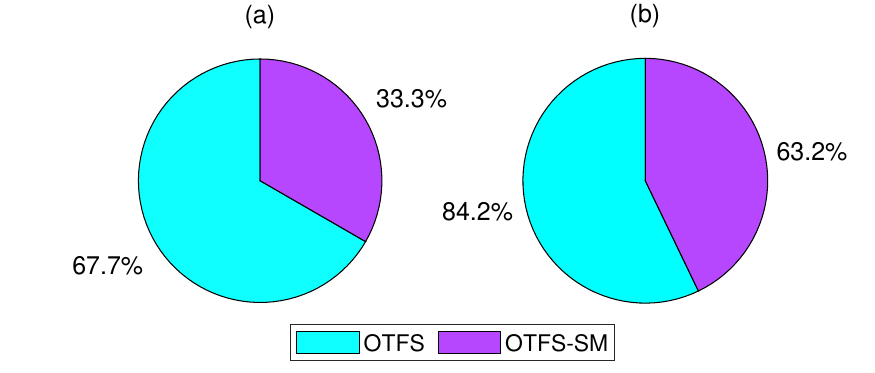}}
	\caption{Energy saving percentages of the proposed OTFS-MBM system compared to OTFS and OTFS-SM systems for (a) $M_q=4$, $N_T=4$, $n_{RF}=4$, $N=2$, $M=2$ and (b) $M_q=8$, $N_T=16$, $n_{RF}=16$, $N=4$, $M=4$.}
	\label{energ} 
\end{figure}

The maximum likelihood (ML) detector of the proposed OTFS-MBM system jointly estimates the transmitted symbol and the index of the active MAP at the transmitter, with its formulation for the OTFS-MBM system for $i^\text{th}$ DD grid given as follows:
\begin{eqnarray}\label{eq11}
 [\hat{q}_i, \hat{\varpi}_i] & =& \text{arg}\underset{q, \varpi}{\mathrm{min}} \ \Big|\Big| \bar{\mathbf{y}} - \textbf{h}_{i,\varpi} x_q \Big|\Big|^2,
\end{eqnarray}
where  $\textbf{h}_{i,\varpi} \in \mathbb{C}^{N_RNM \times 1}$ is the $\varpi^\text{th}$ column vector of $\textbf{H}_{i}$ and $\textbf{H}_{i} \in \mathbb{C}^{N_RNM \times 2^{n_{RF}}}$ is the matrix containing the $i^\text{th}$ column of $\mathbf{H}_{\text{M}}^{\text{eff}}$. The ML detector of the OTFS-MBM system repeats $NM$ times to estimate the selected symbol $x_q$ and active MAP index $\varpi$ for each DD grid. As a result, the estimated $NM$ symbols and active MAP indices are expressed as follows:
\begin{eqnarray}\label{eq15}
\hat{\textbf{q}} = [\hat{q}_1, \hat{q}_2, \cdots, \hat{q}_{NM}],
\end{eqnarray}
\begin{eqnarray}\label{eq16}
\hat{\boldsymbol{\varpi}} = [\hat{\varpi}_1, \hat{\varpi}_2, \cdots, \hat{\varpi}_{NM}].
\end{eqnarray}
Finally, the bits corresponding to the estimated parameters are used to obtain the estimated data bits for the DD domain.

\section{Performance analyses}
This section presents performance analyses, including throughput, spectral efficiency, and energy saving, for the proposed OTFS-MBM system. Comparative performance analyses of the OTFS-MBM, OTFS, and OTFS-SM systems are conducted for various system parameters, emphasizing the advantages of the proposed OTFS-MBM system, particularly in high-mobility and energy-constrained wireless communication scenarios.

\subsection{Throughput Analysis}
Throughput is an important criterion in wireless communications. It refers to the amount of data bits that can be accurately obtained at the receiver out of the total data bits transmitted from the transmitter. The throughput of the proposed OTFS-MBM system is given as follows:
\begin{equation}\label{throughput}
\mathcal{\alpha} = \frac{\big(1 - \text{BER}_{\text{OTFS-MBM}}\big)}{T_s} \eta_{\text{OTFS-MBM}},
\end{equation}
where $\big(1 - \text{BER}_{\text{OTFS-MBM}}\big)$ represents the probability of correct bit detection within the symbol duration $T_s$. In Fig. \ref{thro_mbm}, the proposed OTFS-MBM, OTFS-SM, and OTFS systems are evaluated under two configurations: (a) with parameters $M_q=4$, $N_T=2$, $n_{RF}=2$, $N=2$, $M=2$, and $N_R=2$, and (b) with the same system parameters except for $N_R=6$. As shown in Fig. \ref{thro_mbm}, the proposed OTFS-MBM consistently achieves higher throughput under different parameters, demonstrating its efficiency in high-mobility scenarios.

\subsection{Data Rate Analysis}
Spectral efficiency refers to the number of data bits transmitted over a band-limited channel during a given transmission period. The spectral efficiency of the proposed OTFS-MBM system is compared with that of OTFS and OTFS-SM in Fig. \ref{spec_mbm}. In Fig. \ref{spec_mbm} (a), the proposed OTFS-MBM, OTFS-SM, and OTFS systems are configured with $M_q=4$, $N_T=2$, $n_{RF}=2$, $N=2$, and $M=2$, while Fig. \ref{spec_mbm} (b) uses system parameters $M_q=8$, $N_T=4$, $n_{RF}=4$, $N=2$, and $M=2$. The spectral efficiencies of OTFS and OTFS-SM are defined as $\eta_\text{OTFS} = NM \log_2(M_q)$ and $\eta_\text{OTFS-SM} = NM \log_2(M_q n_T)$, respectively. Fig. \ref{spec_mbm} shows that the proposed OTFS-MBM system has higher spectral efficiency than OTFS-SM and OTFS systems for cases (a) and (b), especially for large-sized system parameters, emphasizing its superior data transfer capability in high-mobility wireless environments.

\subsection{Energy Saving Analysis}
Low energy consumption in wireless communication systems is an important criterion for sustainability. In this context, the proposed OTFS-MBM system presents a new IM-based approach that effectively reduces energy consumption. The OTFS-MBM system consumes significantly less energy than the conventional OTFS system because it transmits most of its data bits over MAP indices.  The energy saving percentage $\beta_{\text{sav}}$ per $\eta_{\text{OTFS-MBM}}$ bits of the OTFS-MBM system is given as follows:
\begin{equation}\label{energy}
\beta_{\text{sav}}  = \Big(1-\frac{\eta_{\text{d}}}{\eta_{\text{OTFS-MBM}}}\Big)E_b\%,
\end{equation}
where $E_b$ denotes the energy per bit and $\eta_{\text{d}}$ represents the spectral efficiency of other schemes, such as OTFS and OTFS-SM. The energy savings percentages of OTFS-MBM compared to OTFS and OTFS-SM are presented in Fig. \ref{energ}. In Fig. \ref{energ} (a), the system parameters are set to $M_q=4$, $N_T=4$, $n_{RF}=4$, $N=2$, and $M=2$, whereas Fig. \ref{energ} (b) considers $M_q=8$, $N_T=16$, $n_{RF}=16$, $N=4$, and $M=4$. For instance, in Fig. \ref{energ} (b), the OTFS-MBM system achieves energy savings of $84.2\%$ and $63.2\%$ compared to OTFS and OTFS-SM, respectively. The results show that the proposed OTFS-MBM system can transmit the same number of data bits with lower energy consumption.

\begin{figure}[t]
\centering{\includegraphics[width=0.5\textwidth]{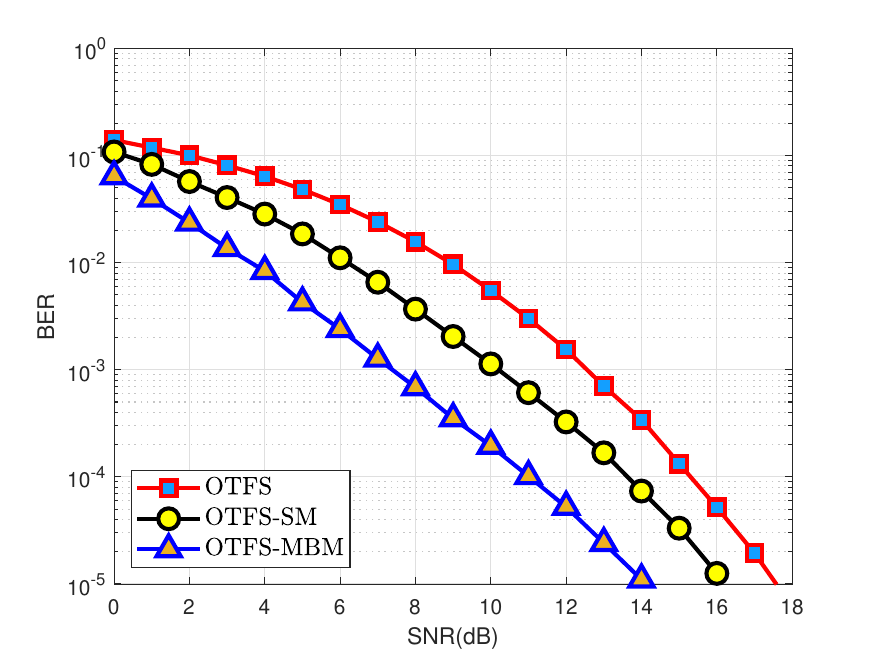}}
	\caption{
    BER performance comparison of the proposed OTFS-MBM, OTFS-SM, and OTFS systems when \(N_R=3\) for $\eta_{OTFS-MBM}=80$ bits.}
	\label{mbm_ber1} 
\end{figure}

\begin{figure}[t]
\centering{\includegraphics[width=0.5\textwidth]{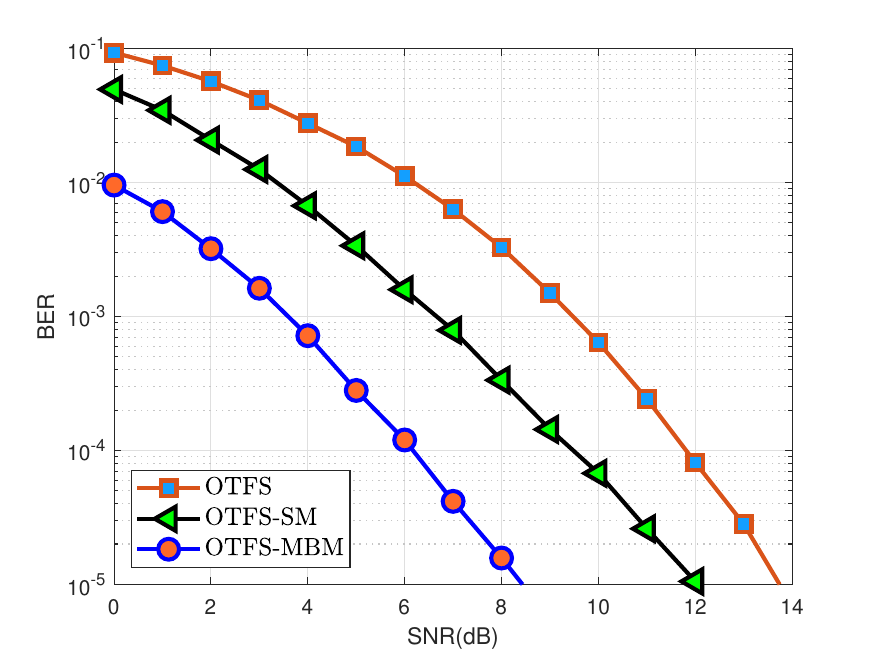}}
	\caption{BER performance comparison of the proposed OTFS-MBM, OTFS-SM, and OTFS systems when \(N_R=5\) for $\eta_{OTFS-MBM}=80$ bits.}
	\label{mbm_ber2} 
\end{figure}

\begin{figure}[t]
\centering{\includegraphics[width=0.5\textwidth]{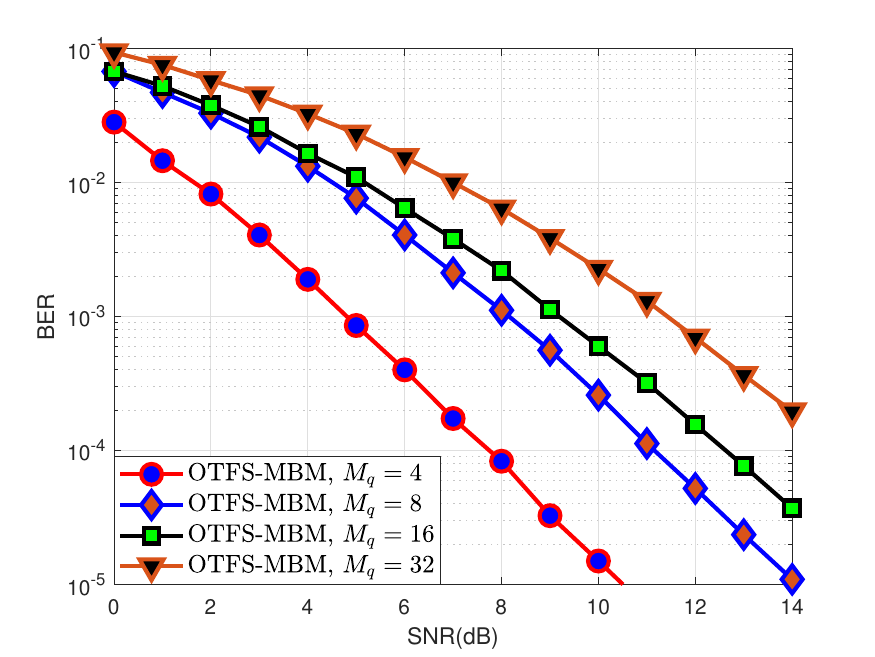}}
	\caption{BER performance comparison of the proposed OTFS-MBM scheme for different $M_q$ values.}
	\label{mbm_mq} 
\end{figure}

\begin{figure}[t]
\centering{\includegraphics[width=0.5\textwidth]{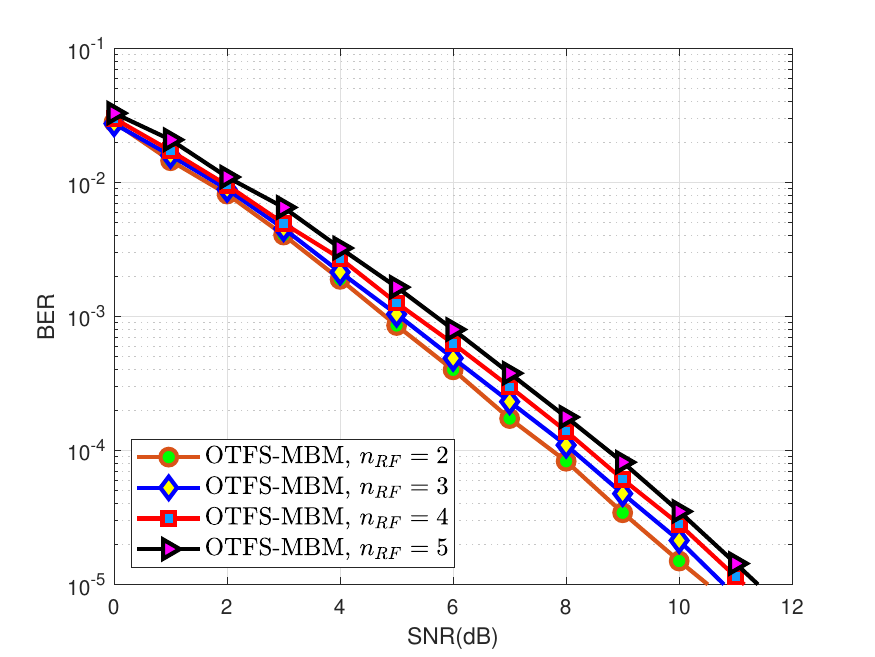}}
	\caption{BER performance comparison of the proposed OTFS-MBM scheme for different $n_{RF}$ values.}
	\label{mbm_mrf} 
\end{figure}


\definecolor{s0}{rgb}{1, 0.85, 1}
\definecolor{s1}{rgb}{1, 0.75, 1}
\definecolor{s2}{rgb}{1, 0.7, 1}
\definecolor{s3}{rgb}{1, 0.65, 1}
\definecolor{s4}{rgb}{1, 0.6, 1}
\definecolor{s5}{rgb}{1, 0.5, 1}




\section{Simulation Results}

This section analyzes the effects of different system parameters on the error performance of the proposed OTFS-MBM system over high-mobility Rayleigh channels. In addition, the performance of the proposed OTFS-MBM system is compared with that of counterpart systems, such as OTFS and OTFS-SM, in the literature. Also, all systems use the QAM technique. In all simulations, the SNR is represented as $\mathrm{SNR (dB)} = 10\log_{10}(E_s/N_0)$.

Fig. \ref{mbm_ber1}  and Fig. \ref{mbm_ber2} \ show the BER performance of OTFS, OTFS-SM, and proposed OTFS-MBM schemes for different numbers of receive antennas. For Fig. \ref{mbm_ber1} and Fig. \ref{mbm_ber2}, \(N_R = 3\) and \(N_R = 5\), respectively.  In Fig. \ref{mbm_ber1}  and Fig. \ref{mbm_ber2}, the proposed OTFS-MBM, OTFS-SM, and OTFS systems are configured with ($M_q=4$, $N_T=1$, $n_{RF}=3$, $N=M=4$); ($M_q=8$, $N_T=4$, $N=M=4$); and ($M_q=32$, $N_T=1$, $N=M=4$), respectively. Fig. \ref{mbm_ber1} and Fig. \ref{mbm_ber2} demonstrate that the proposed OTFS-MBM system achieves superior error performance compared to the OTFS-SM and OTFS systems.



Fig. \ref{mbm_mq} shows the BER performance of the proposed OTFS-MBM for various modulation order \( M_q \) values. The system parameters are  selected  as $M_q=4,8,16,32$, $N_T=1$, $N_R=4$, $n_{RF}=2$, and $N=M=4$. As illustrated in Fig. \ref{mbm_mq}, as \( M_q \) increases, the BER performance of the proposed OTFS-MBM scheme deteriorates. This indicates that higher modulation orders lead to more bit errors at the same SNR. However, the spectral efficiency of the proposed OTFS-MBM system increases as the value of \( M_q \) increases.   


The BER performance of the proposed OTFS-MBM scheme is shown in Fig. \ref{mbm_mrf} for various numbers of RF mirrors ($n_{RF}$). The system parameters are determined  as $ n_{RF}=\{2,3,4,5\}, M_q=4$, $N_T=1$, $N_{R}=4$, and $N=M=4$. It is observed that as \(n_{RF}\) increases, the BER performance of the proposed OTFS-MBM scheme degrades slightly. However, the spectral efficiency of the proposed OTFS-MBM  system increases when \( n_{RF} \) increases.


\section{CONCLUSION}
 This article has proposed an OTFS-based MBM system that has integrated the OTFS technique with MBM to enhance performance and data rates in high-mobility Rayleigh channels. The impact of modulation order, the number of receive antennas, and RF mirrors on the error performance of the OTFS-MBM system has been investigated. It has been demonstrated that increasing the modulation order and the number of RF mirrors has degraded error performance, while a higher number of receive antennas has enhanced it. Comparative analyses have shown that OTFS-MBM has outperformed OTFS and OTFS-SM systems in error performance, throughput, spectral efficiency, and energy efficiency, emphasizing its adaptability to high-mobility wireless environments. These results have shown the proposed OTFS-MBM system as a promising candidate for next-generation wireless networks, offering reliable communication in high-mobility scenarios while efficiently utilizing spectral and energy resources.

\bibliographystyle{IEEEtran}

\bibliography{IEEEabrv,Referanslar}

\end{document}